\documentstyle[preprint,aps,psfig]{revtex}

\input{psfig}
\def\be{\begin{equation}}
\def\ee{\end{equation}}
\def\bea{\begin{eqnarray}}
\def\eea{\end{eqnarray}}

\begin{document}

\title{Can Momentum Correlations Proof Kinetic Equilibration 
       in Heavy Ion Collisions at 160~AGeV?}

\author{M.~Bleicher${}^a$\thanks{E-mail: bleicher@th.physik.uni-frankfurt.de}, 
M. Belkacem${}^a$, C. Ernst${}^a$, H. Weber${}^a$, 
L. Gerland${}^a$, C. Spieles${}^a$, S. A. Bass${}^b$\thanks{Feodor Lynen Fellow of the 
Alexander v. Humboldt Foundation}, H.~St\"ocker${}^a$, W.~Greiner${}^a$}

\address{${}^a$ Institut f\"ur
Theoretische Physik,  J.~W.~Goethe-Universit\"at,\\
D-60054 Frankfurt am Main, Germany}

\address{${}^b$ Department of Physics, Duke University\\
        Durham, N.C. 27708-0305, USA}


\maketitle
\begin{abstract}
We perform an event-by-event analysis of the transverse momentum
distribution of final state particles in central Pb(160AGeV)+Pb
collisions within a microscopic non-equilibrium transport model (UrQMD). 
Strong influence of rescattering is found. 
The extracted momentum distributions show less fluctuations in A+A
collisions than in p+p reactions.
This is in contrast to simplified p+p extrapolations and random walk models.
\end{abstract}

\newpage

The approach to thermalization in relativistic heavy ion collision has gained
a lot of 
attention in the last year. It has been intensively
investigated experimentally and theoretically  \cite{pbm,stachel,cleymans,bass}. 
However, very new event-by-event fluctuation data shed new light onto
this topic. Momentum fluctuations are due to the finite particle number and 
the reaction dynamics. In thermal equilibrium they are
connected to the heat capacity of nuclear matter \cite{landau}.
Additional insight may also be gained by studying particle number fluctuations
which  are connected to another key quantitity, namely the compressibility of
hadronic matter under extreme conditions \cite{mrow}. 

In this letter we perform an event-by-event analysis of central (b$<$3~fm)
Pb+Pb collisions at 160 AGeV within the Ultra-relativistic Quantum
Molecular Dynamics model (UrQMD) \cite{urqmd}. UrQMD is a microscopic
hadronic transport approach based on the covariant propagation of
mesonic and baryonic degrees of freedom. It allows for rescattering and
the formation and decay of resonances and strings. For further details
about the UrQMD model, the reader is referred to Ref. \cite{urqmd}.

\begin{figure}[t]
\vskip 0mm
\vspace{-1.0cm}
\centerline{\psfig{figure=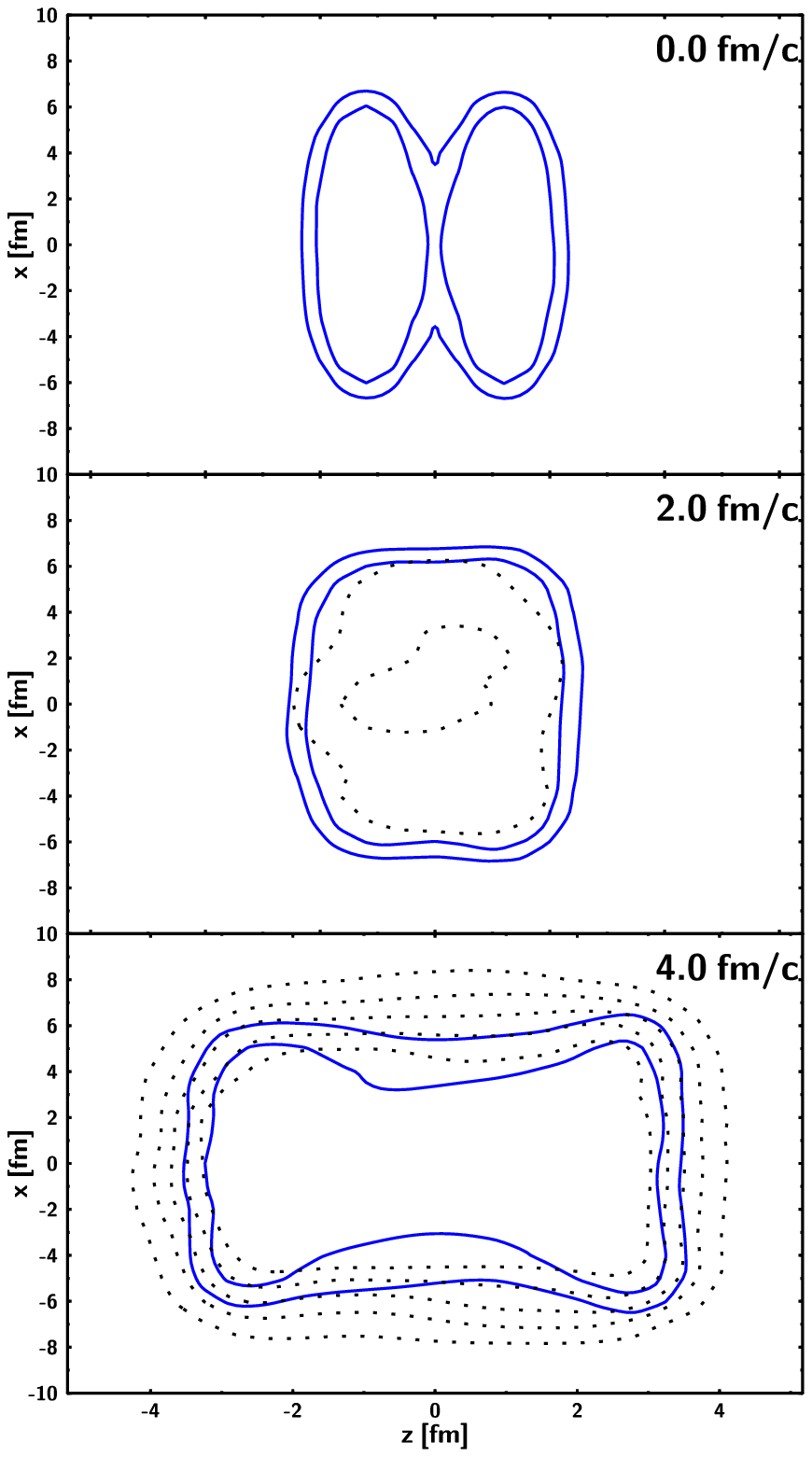,width=10cm}}
\vskip 2mm
\caption{Time evolution of the baryon- (full line) and meson (dotted
line) densities in Pb(160 AGeV)+Pb (b=0~fm) collisions. 
The equal density lines (in the CMS frame) correspond to $\rho_0$, $1.7\rho_0$
for baryons and $0.3\rho_0$, $0.6\rho_0$, $\dots$, $1.5\rho_0$ for the mesons.
The $z$-axis is along the beam axis, $x$ is the impact parameter direction. 
\label{densi}}
\end{figure}

The aim of this investigation is to analyse the fluctuations of
event-by-event distributions both in p+p and nucleus-nucleus reactions.
One of such distributions is the transverse momentum distribution. It
seems obvious that these fluctuations are larger for p+p reactions than
for A+A collisions. In p+p reactions, we have strong correlations in
particle emissions mainly due to conservation laws \cite{kafka}. These
correlations are, however, reduced in A+A collisions due to the larger
amount of energy available in the system and due to rescattering of
secondaries. Indeed, UrQMD calculations show these fluctuations to be
much smaller in Pb+Pb than in p+p, due to the large amount of
meson-baryon (MB) and meson-meson (MM) rescattering. To illustrate this
scenario Fig. \ref{densi} shows the time evolution of the baryon- (full
line) and meson (dotted line) densities in Pb+Pb at 160 AGeV. After
4~fm/c the baryons have complete spatial overlap with the mesons
resulting in high scattering rates with secondaries.

For our purpose, it is convenient to introduce the momentum correlation
parameter $\Phi$ defined as \cite{marek}:
\be
\Phi:=\sqrt{<Z^2>/<N>}-\sqrt{\overline{z^2}}\quad,\quad\mbox{with}\quad
Z=\sum^N_{i=1}z_i \quad,\quad z_i=p_{{\perp_i}}-\overline{p_\perp}   
\label{eq1}
\ee
and $\overline{p_\perp}$ as the particle averaged transverse momentum,
$\overline z$ denotes the particle averaged $z$, $<N>$ the average number of
particles per  event and $<Z>$ is the event averaged $Z$.

Even though this correlation parameter $\Phi$ is not defined in an obvious
way (a careful analysis of this quantity is considered in
\cite{belkacem}) we make use of it to compare our results to recent
investigations \cite{marek-rm,na49}. 

\begin{figure}[t]
\vskip 0mm
\vspace{-1.0cm}
\centerline{\psfig{figure=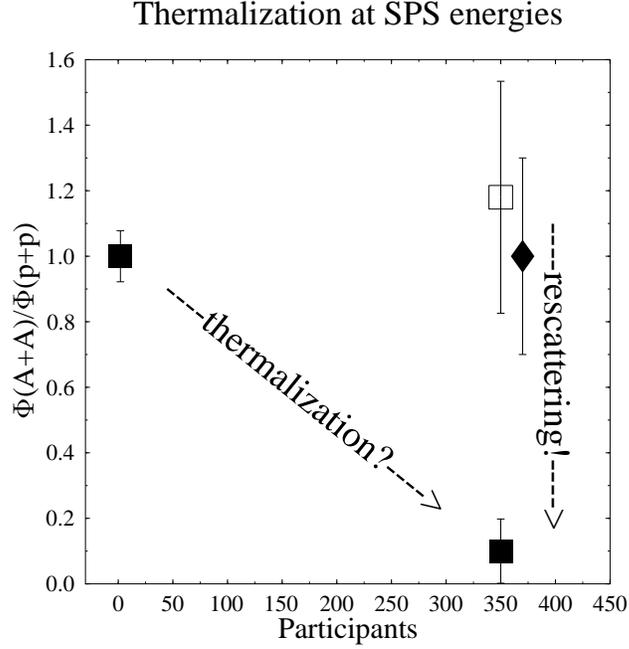,width=10cm}}
\vskip 2mm
\caption{Normalized $\Phi$ values in p+p at 160GeV and Pb+Pb at 160AGeV,
b$<$3~fm. UrQMD calculations with (full squares) and without (empty square) 
rescattering  are compared to FRITIOF calculations (full diamond).
\label{phi}}
\end{figure}

This variable $\Phi$ can provide insight into the emission dynamics of
particles in the course of the collision. The $\Phi$ values can best be 
classified as follows:
\begin{enumerate}
\item $\Phi(A+A)/\Phi(p+p)=1$:\\
                             This indicates that both fluctuations in 
                             p+p and A+A are identical. It corresponds 
                             to a simple extrapolation of proton-proton
                             collisions to describe A+A. This is done 
                             e.g. in the
                             FRITIOF model (see Fig. \ref{phi}).
\item $\Phi(A+A)/\Phi(p+p)<1$:\\
                             The emitted particles in A+A are less
                             correlated than they are in p+p.
                             This is the case in thermal models. 
                             However, UrQMD shows this behaviour 
                             due to the large number
                             of secondary MB and MM collisions  
                             (Fig. \ref{phi})
\item $\Phi(A+A)/\Phi(p+p)>1$:\\
                             The momenta of the particles
                             emitted in  A+A are more strongly correlated 
                             than they are in p+p. This scenario has 
                             recently been investigated in the random walk
                             approach \cite{marek-rm,satz}
\end{enumerate}

The dependence of the correlation variable
$\Phi$ vs. the number of participants normalized to the proton proton
value is shown in Fig. \ref{phi} for the UrQMD model (black squares) while the
FRITIOF calculation is depicted by full diamonds. The full UrQMD
simulation yields a dramatic reduction of the particles momentum correlations
when going from p+p to A+A (indicated by {\em 'thermalization?'}). Thus,
the  UrQMD model 
belongs to the second of the above discussed models. In contrast, a
UrQMD calculation with secondary meson-baryon and meson-meson reactions
turned off - i.e. only baryon-baryon collisions can occur - yields a
$\Phi(A+A)/\Phi(p+p)$ ratio of one. This is shown as the empty square.
Here the $\Phi$
value stays compatible with our expectation from the p+p value. This
observation is supported by FRITIOF calculations (full diamond), which do not
include rescattering of secondaries (see however Ref. \cite{capella}).  
The decrease of the momentum correlation parameter $\Phi$ by one order 
of magnitude when going from p+p to
Pb+Pb (or by the inclusion of rescattering) is in line with NA49 data \cite{na49}.

Does this uncorrelated particle emission indicate global equilibration of the
hadronic source? It has been recently adressed that in each single event
large fluctuations in energy density and isospin happen \cite{spielesprl,bleicherqm}. 
This results in non-thermal initial conditions, which can affect the  final
state of the single event. The information is washed out when one averages over many
reactions. 

\begin{figure}[b]
\vskip 0mm
\vspace{-1.8cm}
\centerline{\psfig{figure=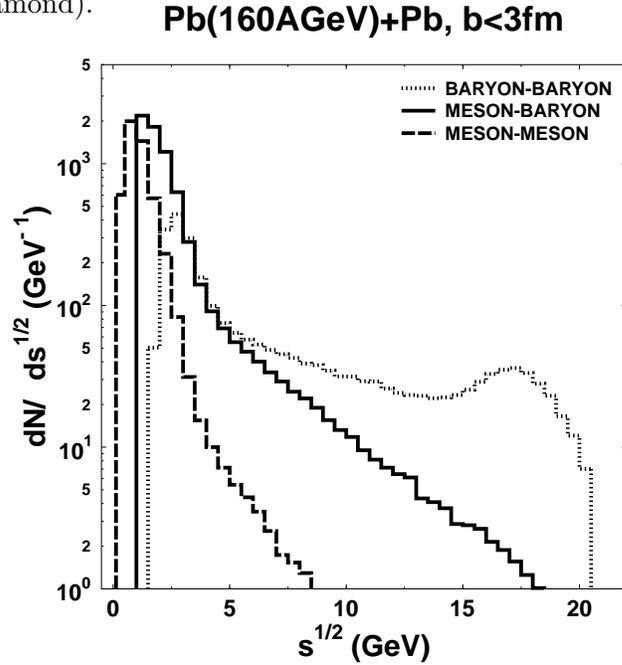,width=10cm}}
\vskip 0mm
\vspace{-.0cm}
\caption{Collision energy spectra of baryon-baryon (dotted line), meson-baryon
(full line) and meson-meson (dashed line) reactions in Pb(160 AGeV)+Pb, b$<$3~fm.
\label{colldyn}}
\end{figure}

In the present model this effect can be pinned down to be due
to the additional MB and MM scattering contributions. In Fig. \ref{colldyn} we have
depicted the collision spectra of all particles in central (b$<$3~fm)
Pb+Pb reactions at 160 AGeV. Overall we find $\approx$1200 baryon-baryon,
$\approx$3400 meson-baryon and $\approx$2500 meson-meson collisions: 
Only 17\% of all
collisions in the Pb+Pb system are due to baryon+baryon extrapolations
in the present model. Thus, the Pb(160AGeV)+Pb system
reveals a far more complex reaction dynamics than collisions at lower
energies or for smaller systems. We propose to study this correlation
ratios for different systems (S+S, p+A, impact parameter dependence in Pb+Pb)
and different beam energies to disentangle these contributions. 
Other quantities should also be introduced to analyse the differences
between correlations in p+p and A+A \cite{belkacem}.
The six times larger number of collisions
compared to the B+B extrapolation (or FRITIOF, resp.) leads to less
correlated momenta of emitted
particles. Therefore, a small $\Phi$ value is a necessary, but not a 
sufficient condition for thermal equilibration.  Thus, from the present momentum 
analysis, on should be careful to draw the conclusion that the
system has thermalized at late times in the collision.

We conclude that heavy ion  collisions at 160~AGeV, exhibit an
intricate interplay between all three collision types (meson-baryon,
meson-meson and baryon-baryon). The most important
interaction for the $p_T$ distribution is meson-baryon scattering with nearly
50\% of all collisions (see also \cite{blei}). This leads to particle
momentum correlation patterns in strong contrast to simplified p+p and p+A 
superpositions/extrapolations. The present model predicts the small $\Phi$
values as measured in Pb(160GeV)+Pb. We would like to draw the attention
of the field to more careful investigation of the hitherto rather unexplored
MM and MB reactions.

\section*{acknowledgements}
This work is supported by the BMBF, GSI, DFG and Graduiertenkolleg 'Schwerionenphysik'. 
M. Bleicher want to thank M. Gazdzicki and G. Roland for stimulating discussions.
M. Bleicher acknowledges support by the Josef Buchmann Foundation.

\newpage

\newpage

\end{document}